\documentclass[aps,prd,groupedaddress,showpacs,twocolumn]{revtex4}

\usepackage[parfill]{parskip}    
\usepackage{graphicx}
\usepackage{amssymb}
\usepackage{epstopdf}
\usepackage{color}
\usepackage{amsmath}

\usepackage{graphicx}
\usepackage{pdfpages}
\usepackage[colorlinks=true,citecolor=blue,urlcolor=magenta,breaklinks]{hyperref}

\begin{document}

\title{Anisotropic neutron stars by gravitational decoupling}
\author{Victor Torres \footnote{victor.torres@yachaytech.edu.ec}}
\address{School of Physical Sciences \& Nanotechnology, Yachay Tech University, 100119 Urcuqu\'i, Ecuador\\}

\author{Ernesto Contreras 
\footnote{econtreras@yachaytech.edu.ec}} 
\address{School of Physical Sciences \& Nanotechnology, Yachay Tech University, 100119 Urcuqu\'i, Ecuador\\}

\begin{abstract}
In this work we obtain an anisotropic neutron star solution by gravitational decoupling starting from
a perfect fluid configuration which has been used to model the compact object PSR J0348+0432. Additionally, we consider the same solution to model the Binary  Pulsar SAX J1808.4-3658 and X-ray Binaries Her X-1 and
Cen X-3 ones. We study the acceptability conditions and obtain that the MGD--deformed solution obey the same physical requirements as its isotropic counterpart. Finally, we conclude that the most stable solutions, according to the adiabatic index and gravitational cracking criterion, are those with the smallest compactness parameters, namely SAX J1808.4-3658 and Her X-1.
\end{abstract}

\maketitle

\section{Introduction}\label{intro}
The first exact interior solution of the Einstein's equations which describe a self--gravitating perfect fluid with constant density 
embedded in a static and spherically symmetric vacuum was obtained by Karl Schwarzschild \cite{sch} and for a long time
isotropic solutions have been broadly considered as suitable interior models of compact objects (see, for example, \cite{lakePF} and references therein). However, very few of these solutions can be considered as physically relevant because they violate some of the elementary conditions that a realistic solution has to satisfy (for a list of physical conditions of interior solutions see, for example, \cite{ivanov}). Even more, in the cases where acceptable solutions
can be found, the perfect fluid model can not be used to deal with situation where it is assumed local anisotropy of pressure, 
that seems to be very reasonable for describing the matter distribution under a variety of circumstances \cite{lemaitre,bowers,cosenza1,cosenza2,herrera92,bondi1,barreto,matrinez,herrera97,herrera98,bondi2,hernandez,herrera2001,aurora,
herreraallstatic,nunez,contrerasAS}.  

In this sense, we may wonder if given an acceptable perfect fluid interior solution, in the sense of \cite{ivanov}, one can be able to extend it to anisotropic domains holding its physical acceptability to some extent. Fortunately, this
question has an affirmative answer. Indeed, the so--called Minimal Geometric Deformation (MGD) method \cite{ovalle2008,
	ovalle2009,ovalle2010,casadio2012,ovalle2013,ovalle2013a,
	casadio2014,casadio2015,ovalle2015,casadio2015b,
	ovalle2016, cavalcanti2016,casadio2016a,ovalle2017,
	rocha2017a,rocha2017b,casadio2017a,ovalle2018,ovalle2018bis,
	estrada2018,ovalle2018a,lasheras2018,gabbanelli2018,sharif2018,sharif2018a,sharif2108b,
	fernandez2018,fernandez2018b,
	contreras2018, estrada, contreras2018a,morales,tello18,
	rincon2018,ovalleplb,contreras2018c,contreras2019,contreras2019a,tello2019,
	contrerasextended,tello2019a,lh2019,estrada2019,gabbanelli2019,ovalle2019a,sudipta2019,linares2019}
have been extensively used to obtain new black hole solutions and to extend interior isotropic models, 
given the number of ingredients which convert it in a versatil and powerful tool to solve the Einstein's equations.

Recently, a perfect fluid interior solution has been reported in Ref. \cite{estevez2018} and has been used to model the neutron star PSR J0348+0432. It is our main goal here extend this solution by MGD to model not only the compact object PSR J0348+0432 
but the Binary Pulsar SAX J1808.4-3658 and X-ray Binaries Her X-1 and Cen X-3 ones. Additionally, we want to explore the physical acceptability  of the anisotropic solution obtained by MGD.

This work is organized as follows. In the next section we review
the main aspects related to the MGD-decoupling method. Section \ref{new} is 
devoted to introduce the new static and spherically symmetric solution and  the anisotropic
solution obtained by MGD is studied in section \ref{anisotropic}. In the last section, we present our final comments  and conclusions.

\section{Einstein Equations and MGD--decoupling}\label{mgd}
This section is devoted to review the main aspects of MGD. To this end, we shall follow closely the original paper on this topic in the context of gravitational decoupling in General Relativity reported in \cite{ovalle2017}. 

As starting point, let us consider the Einstein field equations
\begin{eqnarray}
R_{\mu\nu}-\frac{1}{2}R g_{\mu\nu}=-\kappa^{2}T_{\mu\nu}^{(tot)},
\end{eqnarray}
and assume that the total energy--momentum tensor, $T_{\mu\nu}^{(tot)}$, can be decomposed as
\begin{eqnarray}\label{total}
T_{\mu\nu}^{(tot)}=T_{\mu\nu}^{(m)}+\alpha\theta_{\mu\nu},
\end{eqnarray}
where $T^{(m)}_{\mu\nu}$ is the matter energy momentum for a perfect fluid and $\theta_{\mu\nu}$ an anisotropic source interacting with $T^{(m)}_{\mu\nu}$. Note that,
since the Einstein tensor is divergence free, the total energy momentum tensor $T_{\mu\nu}^{(tot)}$
satisfies
\begin{eqnarray}\label{cons}
\nabla_{\mu}T^{(tot)\mu\nu}=0.
\end{eqnarray}

In what follows, we shall consider spherically symmetric space--times with line element
parametrized as
\begin{eqnarray}\label{le}
ds^{2}=e^{\nu}dt^{2}-e^{\lambda}dr^{2}-r^{2}d\Omega^{2},
\end{eqnarray}
where $\nu$ and $\lambda$ are functions of the radial coordinate $r$ only. 
Now, considering Eq. (\ref{le}) as a solution of the Einstein equations, we obtain
\begin{eqnarray}
\kappa^{2} \tilde{\rho}&=&\frac{1}{r^{2}}+e^{-\lambda}\left(\frac{\lambda'}{r}-\frac{1}{r^{2}}\right)\label{eins1}\\
\kappa^{2} \tilde{p}_{r}&=&-\frac{1}{r^{2}}+e^{-\lambda}\left(\frac{\nu'}{r}+\frac{1}{r^{2}}\right)\label{eins2}\\
\kappa^{2} \tilde{p}_{\perp}&=&\frac{e^{-\lambda}}{4}\left(\nu'^{2}-\nu '\lambda '+2\nu''
+2\frac{\nu'-\lambda'}{r}\right)\label{eins3},
\end{eqnarray}
where the primes denote derivation respect to the radial coordinate and we have defined
\begin{eqnarray}
\tilde{\rho}&=&\rho+\alpha\theta^{0}_{0}\label{rot}\\
\tilde{p}_{r}&=&p-\alpha\theta^{1}_{1}\label{prt}\\
\tilde{p}_{\perp}&=&p-\alpha\theta^{2}_{2}.\label{ppt}
\end{eqnarray} 
Note that, at this point, the decomposition (\ref{total}) seems as a simple separation of the constituents of the matter sector. What is more, given the non--linearity of Einstein's equations, such a decomposition does not lead to a decoupling of two set of equations, one for each source involved. However, contrary to the broadly belief, the decoupling is possible in the context of MGD. The method consists in to introduce a geometric deformation in the the metric functions given by
\begin{eqnarray}
\nu&=&\xi+\alpha g\\
e^{-\lambda}&=&\mu +\alpha f\label{def},
\end{eqnarray}
where $\{g,f\}$ are the so--called decoupling functions and $\alpha$ is a free parameter that
``controls'' the deformation. It is worth mentioning that although a general treatment considering deformation in both components of the metric is possible (see Ref. \cite{ovalleplb}), in this work we shall concentrate in the particular case $g=0$ and $f\ne 0$.  Doing so, we obtain
two sets of differential equations: one describing an isotropic system sourced by
the conserved energy--momentum tensor of a perfect fluid $T^{(m)}_{\mu\nu}$ and the other
set corresponding to quasi--Einstein field equations sourced by $\theta_{\mu\nu}$. More precisely, 
we obtain
\begin{eqnarray}
\kappa ^2 \rho &=&\frac{1-r \mu'-\mu}{r^{2}}\label{iso1}\\
\kappa ^2 p&=&\frac{r \mu  \nu '+\mu -1}{r^{2}}\label{iso2}\\
\kappa ^2 p&=&\frac{\mu ' \left(r \nu '+2\right)+\mu  \left(2 r \nu ''
+r \nu '^2+2 \nu '\right)}{4 r}\label{iso3},
\end{eqnarray}
with
\begin{eqnarray}
\nabla_{\mu}T^{(m)\mu\nu}=p'
-\frac{\nu'}{2}(\rho+p)=0,
\end{eqnarray}{}
for the perfect fluid
and
\begin{eqnarray}
\kappa ^2  \theta^{0}_{0}&=&-\frac{r f'+f}{r^{2}}\label{aniso1}\\
\kappa ^2 \theta^{1}_{1}&=&-\frac{r f \nu '+f}{r^{2}}\label{aniso2}\\
\kappa ^2\theta^{2}_{2}&=&-\frac{f' \left(r \nu '+2\right)+f \left(2 r \nu ''+r \nu '^2+2 \nu '\right)}{4 r}\label{aniso3},
\end{eqnarray}
for the source $\theta_{\mu\nu}$ that, whenever 
$\theta^{1}_{1}\ne\theta^{2}_{2}$, induce local anisotropy in the system. It is worth noticing that the conservation equation $\nabla_{\mu}\theta^{\mu}_{\nu}=0$ leads to
\begin{eqnarray}
(\theta^{1}_{1})'-\frac{\nu'}{2}(\theta^{0}_{0}-\theta^{1}_{1})-\frac{2}{r}(\theta^{2}_{2}-\theta^{1}_{1})=0.
\end{eqnarray}
which is a linear combination of Eqs. (\ref{aniso1}), (\ref{aniso2}) and (\ref{aniso3}). In this
sense, there is no exchange of energy--momentum tensor between the perfect fluid and the
anisotropic  source and henceforth interaction is purely gravitational.\\
As explained in previous sections, the MGD has been successfully used to extend
isotropic solutions to anisotropic domains. More precisely, given a metric functions $\{\nu,\mu\}$ 
sourced by a perefect fluid $\{\rho,p\}$ that solve Eqs. (\ref{iso1}), (\ref{iso2}) and (\ref{iso3}),
the deformation function $f$ can be found from Eqs. (\ref{aniso1}), (\ref{aniso2}) and (\ref{aniso3})
after choosing suitable conditions on the anisotropic source $\theta_{\mu\nu}$. It is worth mentioning
that the case we are dealing with demands for an exterior Schwarzschild solution. In this case, the
matching condition leads to the extra information required to completely solve the system. 

Defining $\mu(r)=1+\frac{m(r)}{r}$ in (\ref{def}) the interior solution parametrized with
(\ref{le}) reads
\begin{eqnarray}
ds^{2}=e^{\nu}dt^{2}-\left(1-\frac{2m(r)}{r}+\alpha f\right)^{-1}dr^{2}-r^{2}d\Omega^{2}.
\end{eqnarray}
Now, outside of the distribution the space--time is that of Schwarzschild, given by
\begin{eqnarray}
ds^{2}=\left(1-\frac{2M}{r}\right)dt^{2}-\left(1-\frac{2M}{r}\right)dr^{2}-r^{2}d\Omega^{2}.
\end{eqnarray}
In order to match smoothly the two metrics above on the
boundary surface $\Sigma$, we must require the continuity of
the first and the second fundamental form across that
surface. Then it follows
\begin{eqnarray}
e^{\nu_{\Sigma}}=1-\frac{2 M}{r_{\Sigma}}\\
e^{\lambda_{\Sigma}}=1-\frac{2 M}{r_{\Sigma}}\\
\tilde{p}_{r_{\Sigma}}=0.
\end{eqnarray}
Note that, the condition on the radial pressure leads to
\begin{eqnarray}\label{cond}
p(r_{\Sigma})-\alpha\theta^{1}_{1}(r_{\Sigma})=0.
\end{eqnarray}
Regardingly, if the original perfect fluid match smoothly with the Schwarzschild solution, i.e, 
$p(r_{\Sigma})=0$, Eq. (\ref{cond}) can be satisfied by demanding $\theta^{1}_{1}\sim p$. Of course,
the simpler way to satisfy the requirement on the radial pressure is assuming the so--called mimic constraint \cite{ovalle2017} for the pressure, namely
\begin{eqnarray}\label{mimetic}
\theta^{1}_{1}= p,
\end{eqnarray}
in the interior of the star.
At this point a couple of comments are in order. First, it is remarkable that assuming the mimic constraint the continuity of the second fundamental is straightforward.
Second, the mimic constrain leads to an simple algebraic equation to solve for the decoupling function $f$ after replacing (\ref{mimetic}) in (\ref{aniso2}). In this sense, the only remaining step to construct an anisotropic interior solution following MGD is to choose a suitable perfect fluid as a seed to solve the system 
(\ref{aniso1}), (\ref{aniso2}) and (\ref{aniso3})
via the mimic constrain to finally obtain the interior solution with local anisotropies described by $\{\nu,\lambda,\tilde{\rho},\tilde{p}_{r},\tilde{p}_{\perp}\}$

\section{Perfect fluid neutron star}\label{new}
In this section we briefly review some aspects about the isotropic spherically symmetric solution reported in Ref. \cite{estevez2018}.

The line element of the interior of the star reads
\begin{eqnarray}\label{metricnew}
ds^{2}=y^{2}(r)dt^{2}-\frac{dr^{2}}{B(r)}-r^{2}(d\theta^{2}+sin\theta^{2}d\phi^{2}),
\end{eqnarray}
where
\begin{align}
y(r)&=\frac{C(5+4a r^{2})}{(1+a r^{2})^{1/2}}\\
B(r)&=\frac{(5+11 a r^{2}+6a^{2}r^{4}-4ar^{2}S(r))(1+a r^{2})}{5+12 ar^{2}+8a^{2}r^{4}},
\end{align}
with
\begin{equation}
S(r)=\frac{(1+ar^{2})^{2}(A+ arctanh(\frac{1+2 a r^{2}}{(5+12 ar^{2}+8a^{2}r^{4})^{1/2}}))}{(5+12 ar^{2}+8a^{2}r^{4})^{1/2}}.
\end{equation}
where $C$, $a$ and $A$ are constants. The metric (\ref{metricnew}) is
a solution of the Einstein Field equation $R_{\mu\nu}-\frac{1}{2}g_{\mu\nu}R=-\kappa^{2}T^{(m)}_{\mu\nu}$
where the isotropic source reads
\begin{eqnarray}
\kappa^{2}\rho &=&-\frac{a(60+256ar^{2}+486a^{2}r^{4}+42a^{3}r^{6}+144a^{4}r^{8})}{(5+12ar^{2}+8a^{2}r^{4})^{2}}\nonumber\\
                          &&+\frac{12a(5+15ar^{2}+16a^{2}r^{4}+8a^{3}r^{6})S(r)}{(5+12ar^{2}+8a^{2}r^{4})^{2}}\\
\kappa^{2}p&=&\frac{a(50+167ar^{2}+190a^{2}r^{4}+72a^{3}r^{6})}{(5+12ar^{2}+8a^{2}r^{4})(5+4ar^{2})}\nonumber\\
                  &&-\frac{4a(5+15ar^{2}+12a^{2}r^{4})S(r)}{(5+12ar^{2}+8a^{2}r^{4})(5+4ar^{2})}.
\end{eqnarray}
A it is shown in \cite{estevez2018}, there is a  convenient parametrization which allows to rewrite the solution as a function of dimensionless quantities given by
\begin{eqnarray}
r &\to& x R\\
aR^{2} &\to& \omega,
\end{eqnarray}
where $R$ is the radius of the star. Furthermore, the normalized radius, $x\in(0,1)$, and the parameter $\omega$ are dimensionless quantities. It is worth noticing that, on one hand, the parameterization allows to write the quantities $\kappa^{2}R^{2}\rho$ and $\kappa^{2}R^{2}p$ as functions of $x$ and $\omega$ only.
On the other hand, the matching conditions lead to (see \cite{estevez2018} for details)
\begin{eqnarray}
A&=&\frac{(50+16\omega +197 \omega^{2}+72\omega^{3})\sqrt{5+12\omega+8\omega^{2}}}{4(5+15\omega+12\omega^{2})(1+\omega)^{2}}\nonumber\\
&&-arctanh\left(\frac{1+2\omega}{\sqrt{5+12\omega+8\omega^{2}}}\right),
\end{eqnarray}
from where we obtain a relationship between $\omega$ and the compacteness $u$ given by
\begin{eqnarray}\label{rel}
u=\frac{\omega(3+4\omega)}{5+15\omega+12\omega^{2}}.
\end{eqnarray}
Note that, Eq. (\ref{rel}) allows to set the
the free parameter $\omega$ once the compactness $u$ is fixed. For example, in Ref. \cite{estevez2018} the authors fixed the compactness parameter using data associated to PSR J0348+0432.

\section{Anisotropic neutron star by MGD}\label{anisotropic}
In what follows, we shall obtain the anisotropic solution implementing the mimic constraint
studied in the previous section. More precisely, from Eq. (\ref{mimetic}), we obtain
\begin{eqnarray}\label{f}
f&=&-\frac{x^2 \omega  \left(x^2 \omega +1\right) \left(x^2 \omega  \left(2 x^2 \omega  \left(36 x^2 \omega +95\right)+167\right)+50\right)}{\left(4 x^2 \omega  \left(2 x^2 \omega +3\right)+5\right) \left(3 x^2 \omega  \left(4 x^2 \omega +5\right)+5\right)}\nonumber\\
&&+\frac{4 x^2 \omega  \left(x^2 \omega +1\right) \left(3 x^2 \omega  \left(4 x^2 \omega +5\right)+5\right) S(x;\omega )}{\left(4 x^2 \omega  \left(2 x^2 \omega +3\right)+5\right) \left(3 x^2 \omega  \left(4 x^2 \omega +5\right)+5\right)},
\end{eqnarray}
where we have used the parametrization introduced in the previous section. 

Note that, for fixed  $(u,\omega)$
the decoupling function, $f$, is a function of $x$ only. However, it is worth noticing that
after implementing MGD the metric function $\lambda$ given by Eq. (\ref{def}) and the matter sector
of the total solution, namely, $\tilde{\rho}$, $\tilde{p}_{r}$ and 
$\tilde{p}_{\perp}$ which are obtained from Eqs. (\ref{def}), (\ref{aniso1}), (\ref{aniso2}) and (\ref{aniso3}) respectively, will depend on the free parameter $\alpha$.

The next step in the program consist in to calculate $\{\lambda,\tilde{\rho},\tilde{p}_{r},\tilde{p}_{\perp}\}$. However, as can be checked by the reader, the resulting functions are too long
that their explicit form is not illuminating at all. In this sense, we shall dedicate the rest of the manuscript in to illustrate their properties by graphical analysis. Furthermore, we shall fix the compactness parameter taking values from table \ref{t1} (see \cite{tello2019} and references therein)
\begin{table}
\begin{tabular}{cccc}
\hline
\textbf{\begin{tabular}[c]{@{}c@{}}Compact Star \\ Model\end{tabular}} & \textbf{\begin{tabular}[c]{@{}c@{}}Mass \\ in $M_{\odot}$\end{tabular}} & \textbf{\begin{tabular}[c]{@{}c@{}}Radius $R$ \\ in Km\end{tabular}} & \textbf{\begin{tabular}[c]{@{}c@{}}Compactness \\ factor $u$\end{tabular}} \\ \hline
SAX J1808.4-3658                                                       & 0.85                                                                    & 9.5                                                                  & 0.13                                                                       \\
Her X-1                                                                & 0.9                                                                     & 8.1                                                                  & 0.15478                                                                    \\
Cen X-3                                                                & 1.49                                                                    & 10.8                                                                 & 0.2035                                                                     \\
PSR J0348+0432.                                                        & 1.97                                                                    & 12.957                                                               & 0.229365                                                                   \\ \hline
\end{tabular}
\caption{\label{t1} Compactness parameter for some compact stars}	
\end{table}
\section{Conditions for physical viability of interior solutions} 
The study of acceptability
conditions of interior solutions is important because, as it is well known, the goal is not only to solve Einstein's equations but to demonstrate that they are suitable to describe a physical system. In this section we explore some of these conditions.

\subsection{Matter sector}
\begin{figure}[h!]
\centering
\includegraphics[scale=0.52]{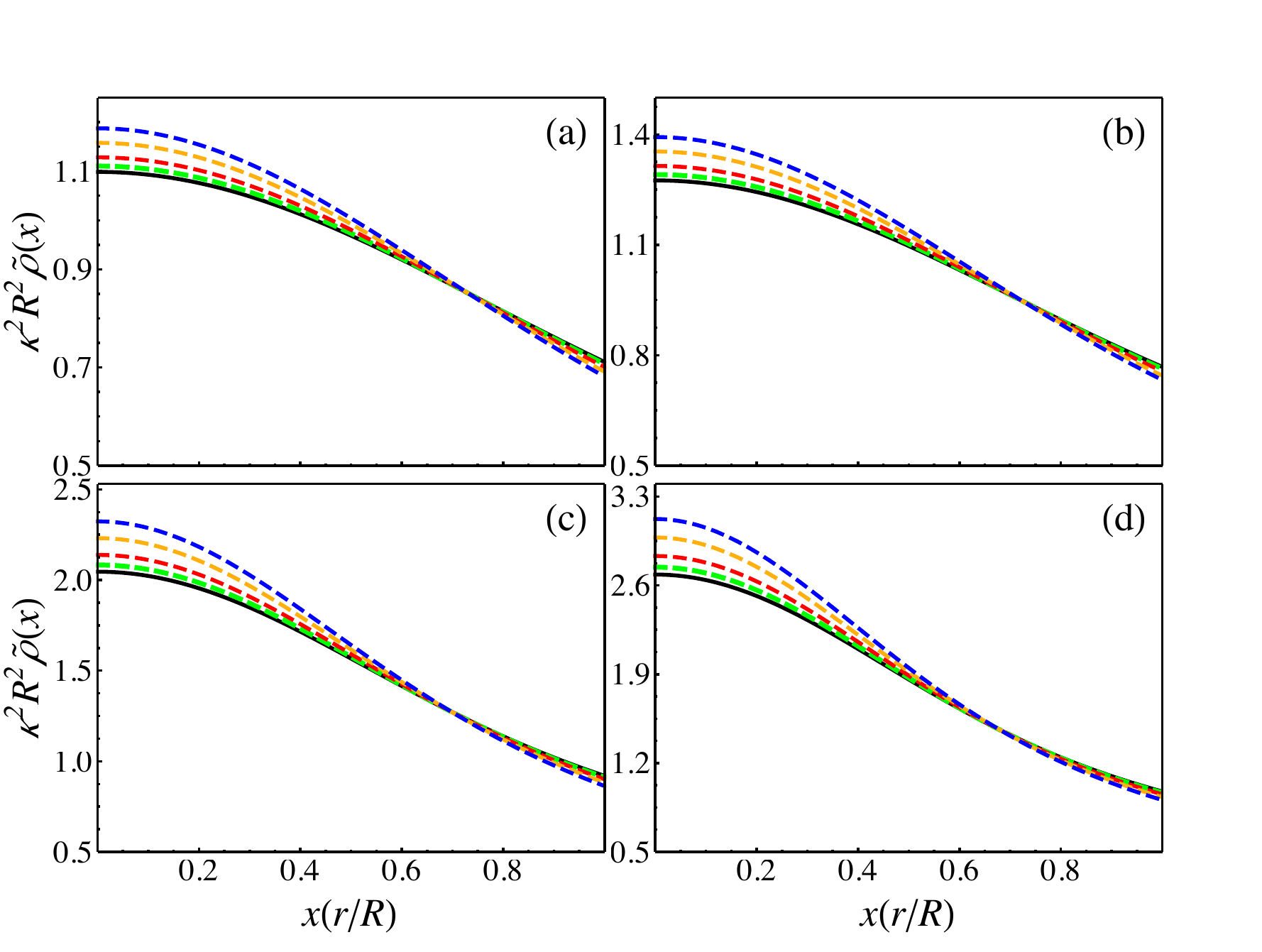}
\caption{\label{density} 
Effective density $\tilde{\rho}$ for a) $u=0.13$, b) $u=0.15478$, c) $u=0.2035$, d) $u=0.229365$.  $\alpha=0$ (black solid line), $\alpha=0.04$, (green dashed line),
$\alpha=0.1$ (red dashed line), $\alpha=0.2$ (yellow dashed line),
$\alpha=0.3$ (blue dashed line)
}
\end{figure}

\begin{figure}[h!]
\centering
\includegraphics[scale=0.52]{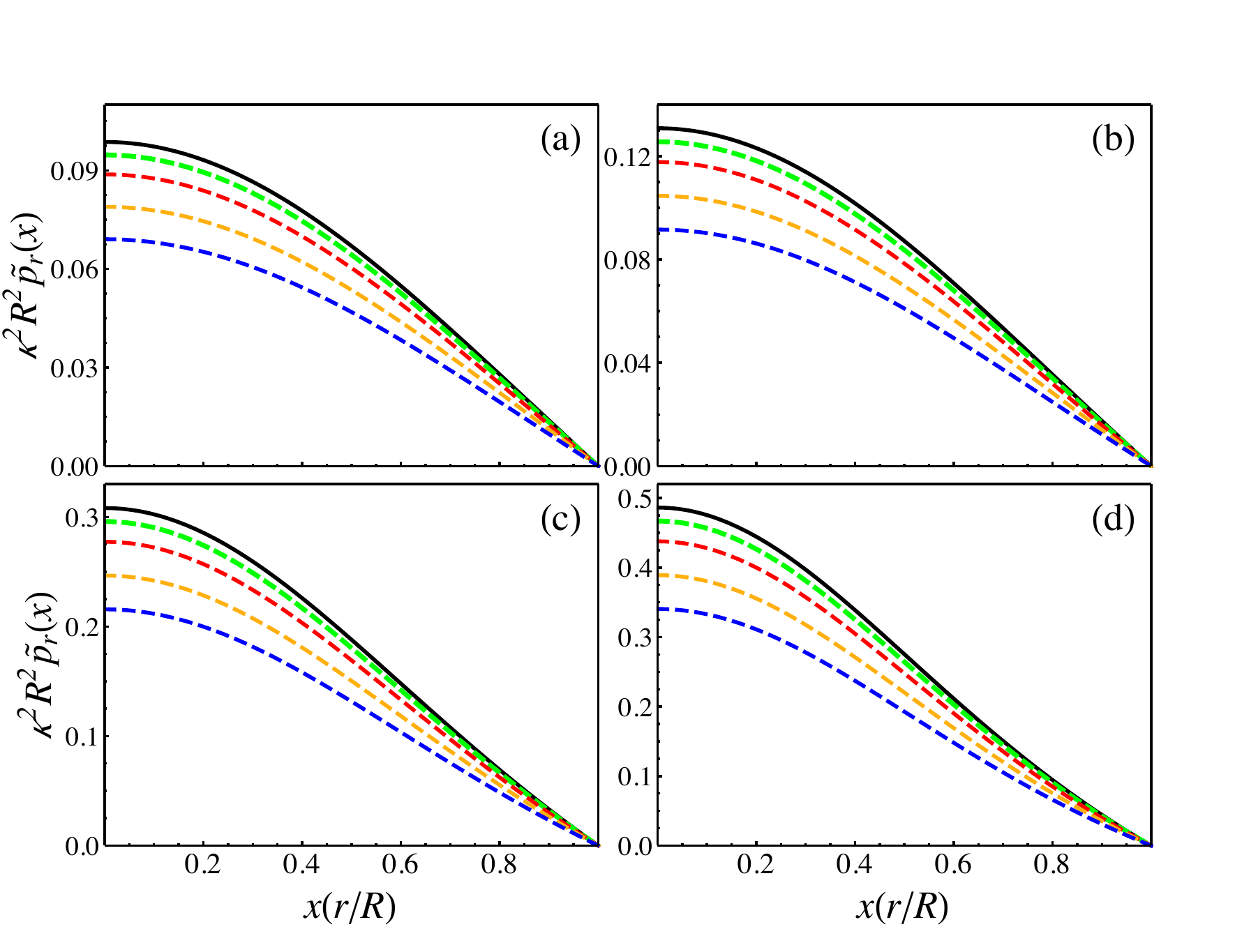}
\caption{\label{radpress} 
Effective radial pressure $\tilde{p}_{r}$. For 
a) $u=0.13$, b) $u=0.15478$, c) $u=0.2035$, d) $u=0.229365$.   $\alpha=0$ (black solid line), $\alpha=0.04$, (green dashed line),
$\alpha=0.1$ (red dashed line), $\alpha=0.2$ (yellow dashed line),
$\alpha=0.3$ (blue dashed line).
}
\end{figure}

\begin{figure}[h!]
\centering
\includegraphics[scale=0.52]{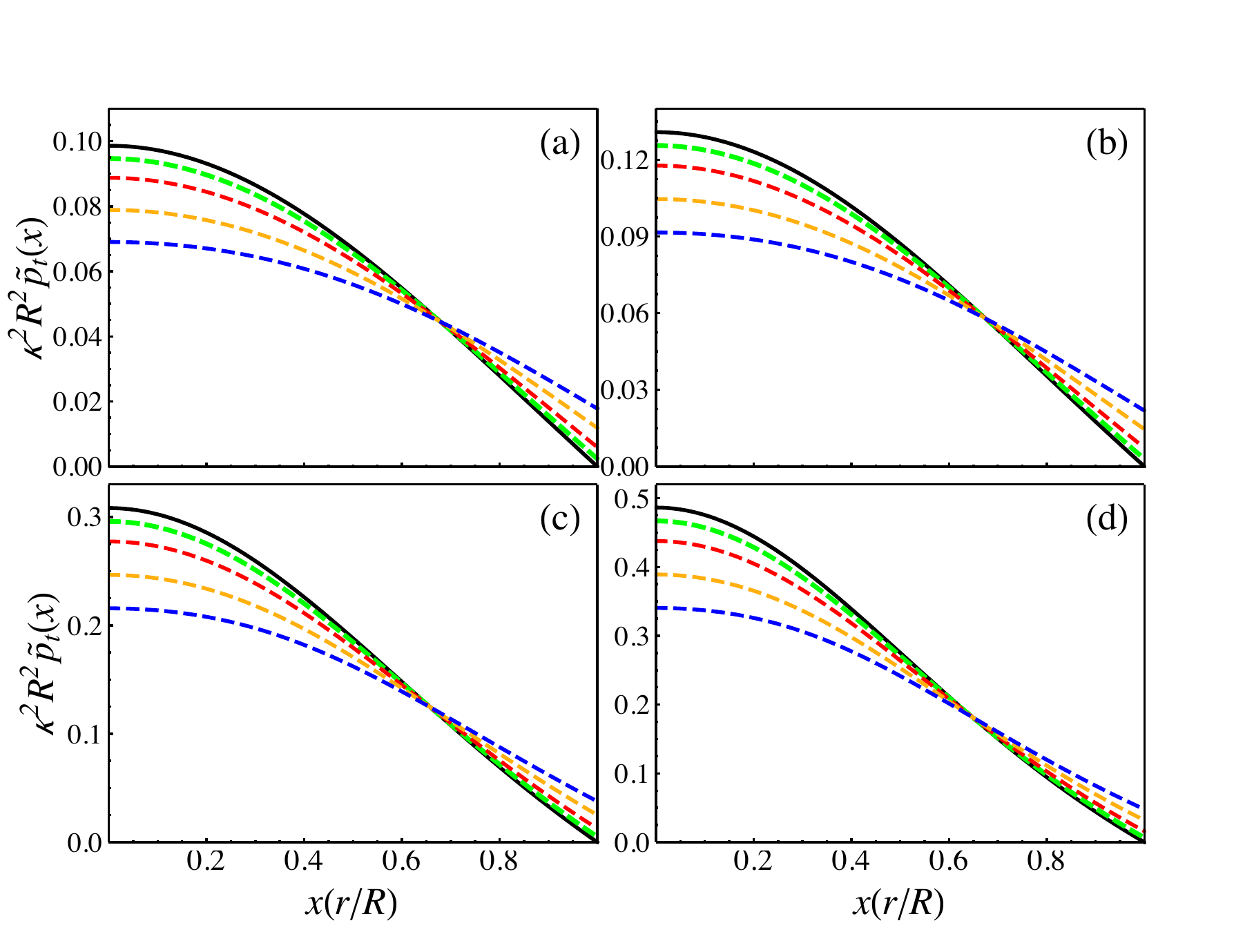}
\caption{\label{perppress} 
Effective tangential pressure $\tilde{p}_{t}$. For 
a) $u=0.13$, b) $u=0.15478$, c) $u=0.2035$, d) $u=0.229365$.  $\alpha=0$ (black solid line), $\alpha=0.04$, (green dashed line),
$\alpha=0.1$ (red dashed line), $\alpha=0.2$ (yellow dashed line),
$\alpha=0.3$ (blue dashed line).
}
\end{figure}

A physically acceptable 
interior solution should have
positive  densities and pressures. Moreover, the density and pressures have to reach a maximum at the center and
decrease monotonously toward the surface with $\tilde{p}_{r}\ge \tilde{p}_{\perp}$. The profiles of $\tilde{\rho}$, $\tilde{p}_{r}$ and $\tilde{p}_{\perp}$ are shown in 
in figures \ref{density}, \ref{radpress}, \ref{perppress} respectively

Note that all the profiles satisfy the first requirement about positivity
and montonouly decreasing for all the values of $\alpha$. Interestingly, as $\alpha$ increases, the density at 
the center becomes greater in contrast to the behaviour on the surface. On the contrary, the radial and the traverse pressure decrease when $\alpha$ grows.

\begin{figure}[h!]
\centering
\includegraphics[scale=0.52]{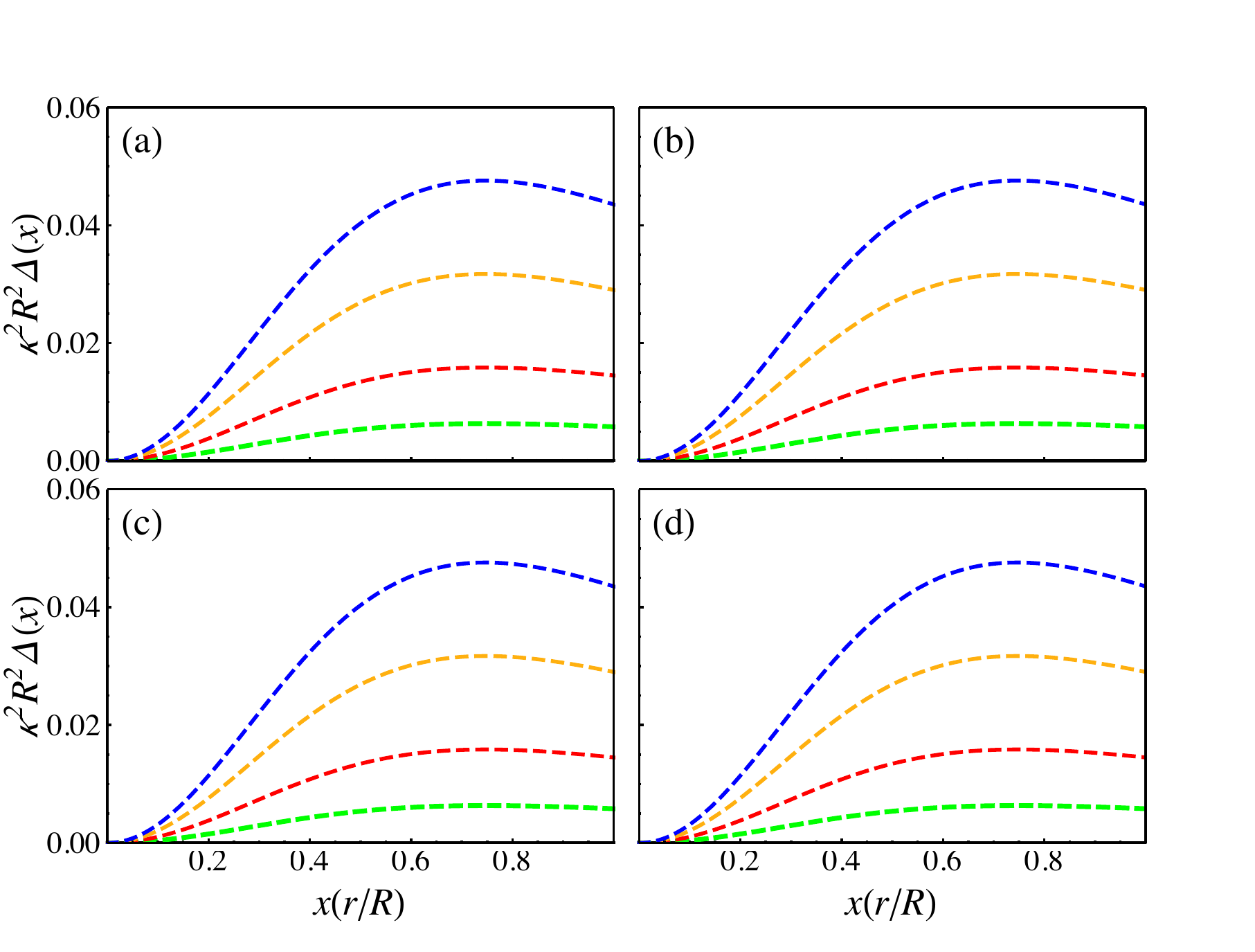}
\caption{\label{anisotropy} 
Anisotropic factor , for 
a) $u=0.13$, b) $u=0.15478$, c) $u=0.2035$, d) $u=0.229365$.  $\alpha=0$ (black solid line), $\alpha=0.04$, (green dashed line),
$\alpha=0.1$ (red dashed line), $\alpha=0.2$ (yellow dashed line),
$\alpha=0.3$ (blue dashed line).
}
\end{figure}

For a physically realistic solution is it also required that the radial and the traverse pressure
be the same at the center but  $\tilde{p}_{\perp}>\tilde{p}_{r}$ as we approach to the surface of the compact object. This behaviour is illustrated in fig. \ref{anisotropy} where we have plotted the anisotropy $\Delta=\tilde{p_{\perp}}-\tilde{p}_{r}$

\subsection{Energy conditions}
\begin{figure}[h!]
	\centering
	\includegraphics[scale=0.52]{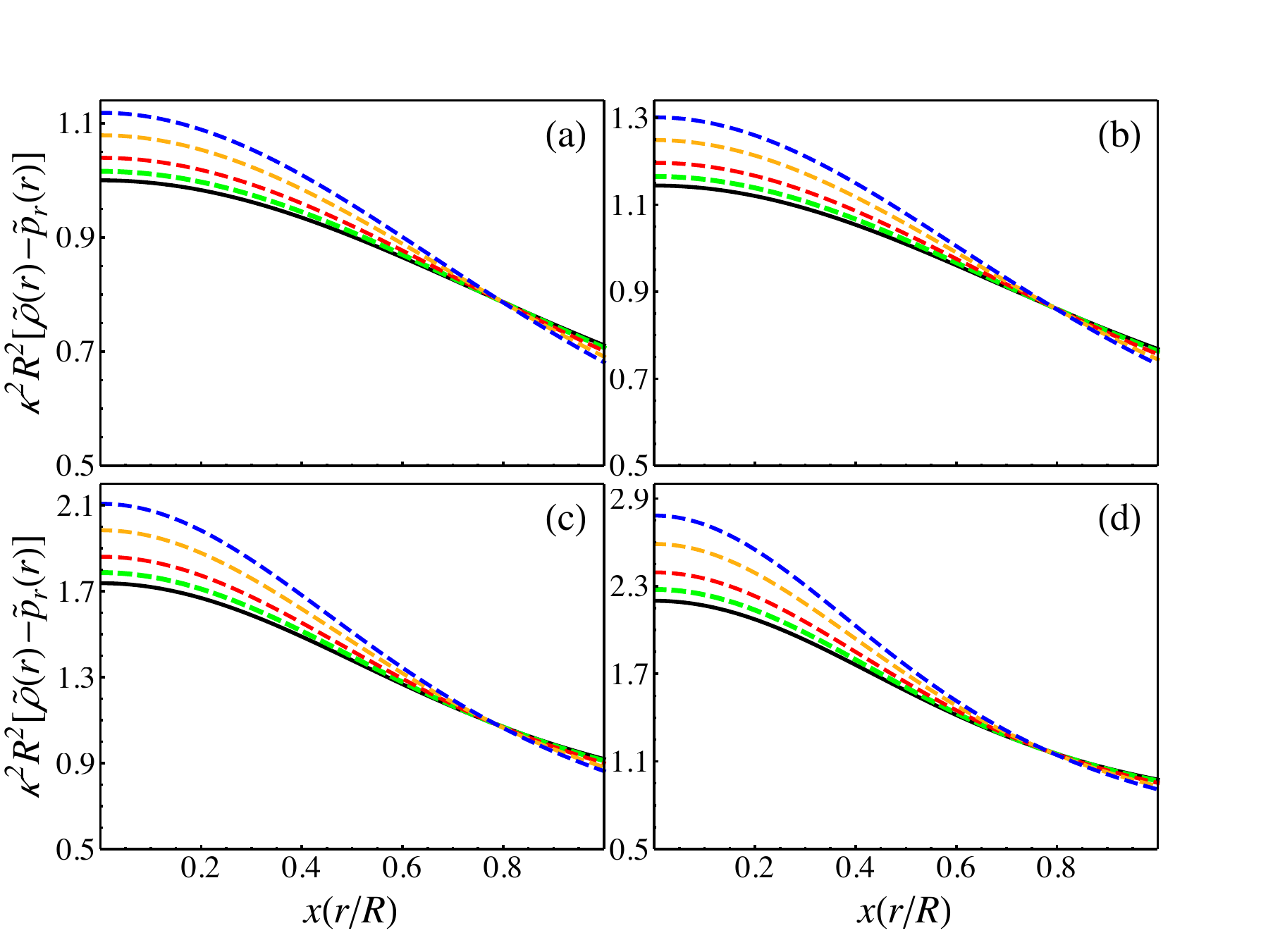}
	\caption{\label{dec1} 
		Dominant energy condition for the radial components, for 
		a) $u=0.13$, b) $u=0.15478$, c) $u=0.2035$, d) $u=0.229365$.  $\alpha=0$ (black solid line), $\alpha=0.04$, (green dashed line),
		$\alpha=0.1$ (red dashed line), $\alpha=0.2$ (yellow dashed line),
		$\alpha=0.3$ (blue dashed line).
	}
\end{figure}

Energy conditions are important in the analysis of interior solutions. 
Accordingly, the dominant energy condition (DEC), which implies that  the  speed of energy flow of matter is less than the speed of light for any observer, should be satisfied by the solution. In order to the DEC be fulfilled, the matter content must
satisfy
\begin{eqnarray}
\tilde{\rho}-\tilde{p_{r}}&\ge& 0\\
\tilde{\rho}-\tilde{p_{\perp}}&\ge& 0.
\end{eqnarray}
As illustrated in figure \ref{dec1}, the DEC is satisfied by all the models for all the values of $\alpha$ under consideration.

\begin{figure}[h!]
\centering
\includegraphics[scale=0.52]{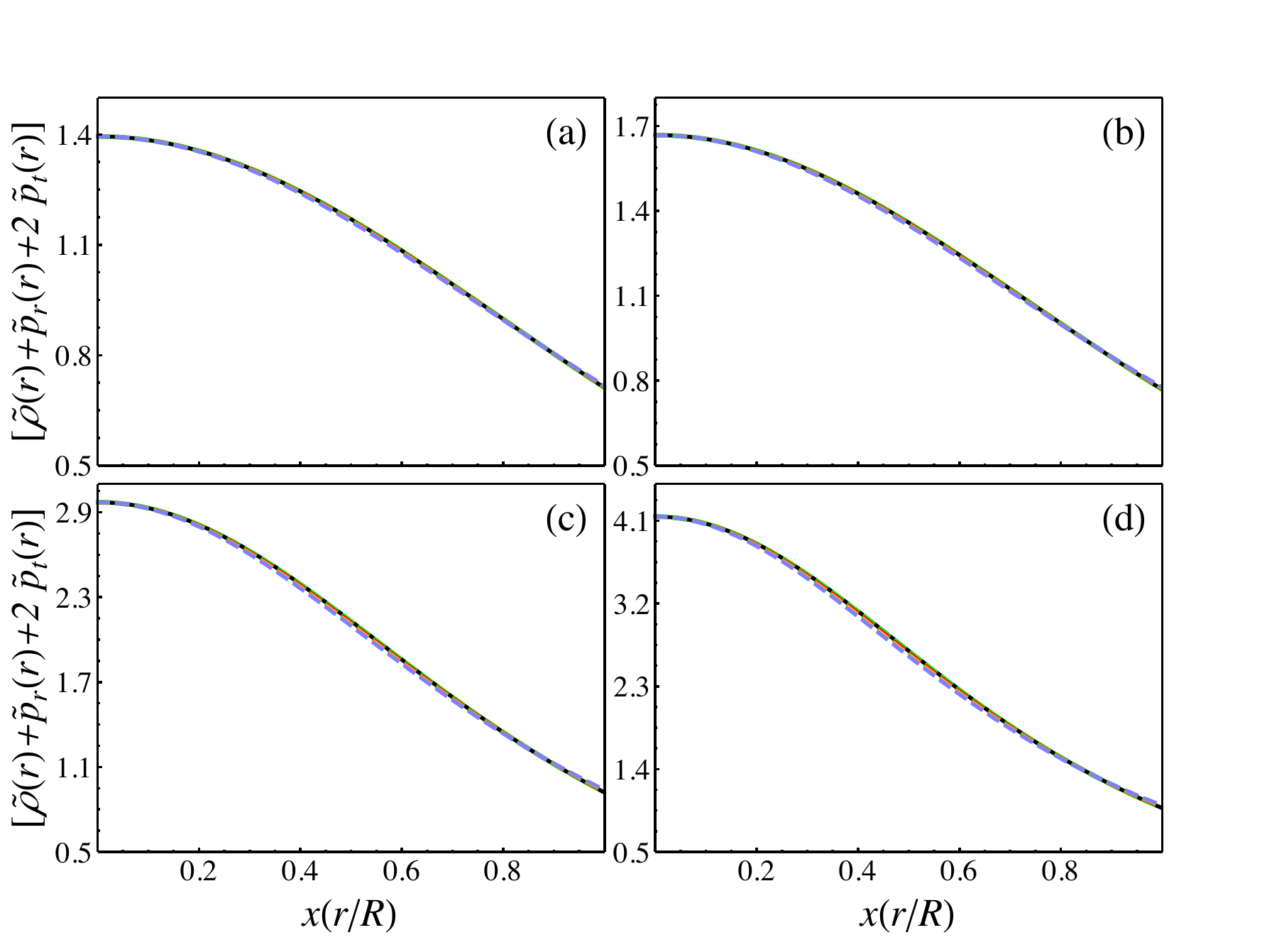}
\caption{\label{strong} 
Strong energy condition, for 
a) $u=0.13$, b) $u=0.15478$, c) $u=0.2035$, d) $u=0.229365$.   $\alpha=0$ (black solid line), $\alpha=0.04$, (green dashed line),
$\alpha=0.1$ (red dashed line), $\alpha=0.2$ (yellow dashed line),
$\alpha=0.3$ (blue dashed line).
}
\end{figure}

It is also desirable that the model satisfies also the strong energy condition (SEC), namely
\begin{eqnarray}
\rho+\sum\limits_{i}p_{i}\ge 0,
\end{eqnarray}
As can be seen in fig. \ref{strong}, this requirment is satisfied in the cases under study.

\subsection{Causality}
\begin{figure}[h!]
\centering
\includegraphics[scale=0.52]{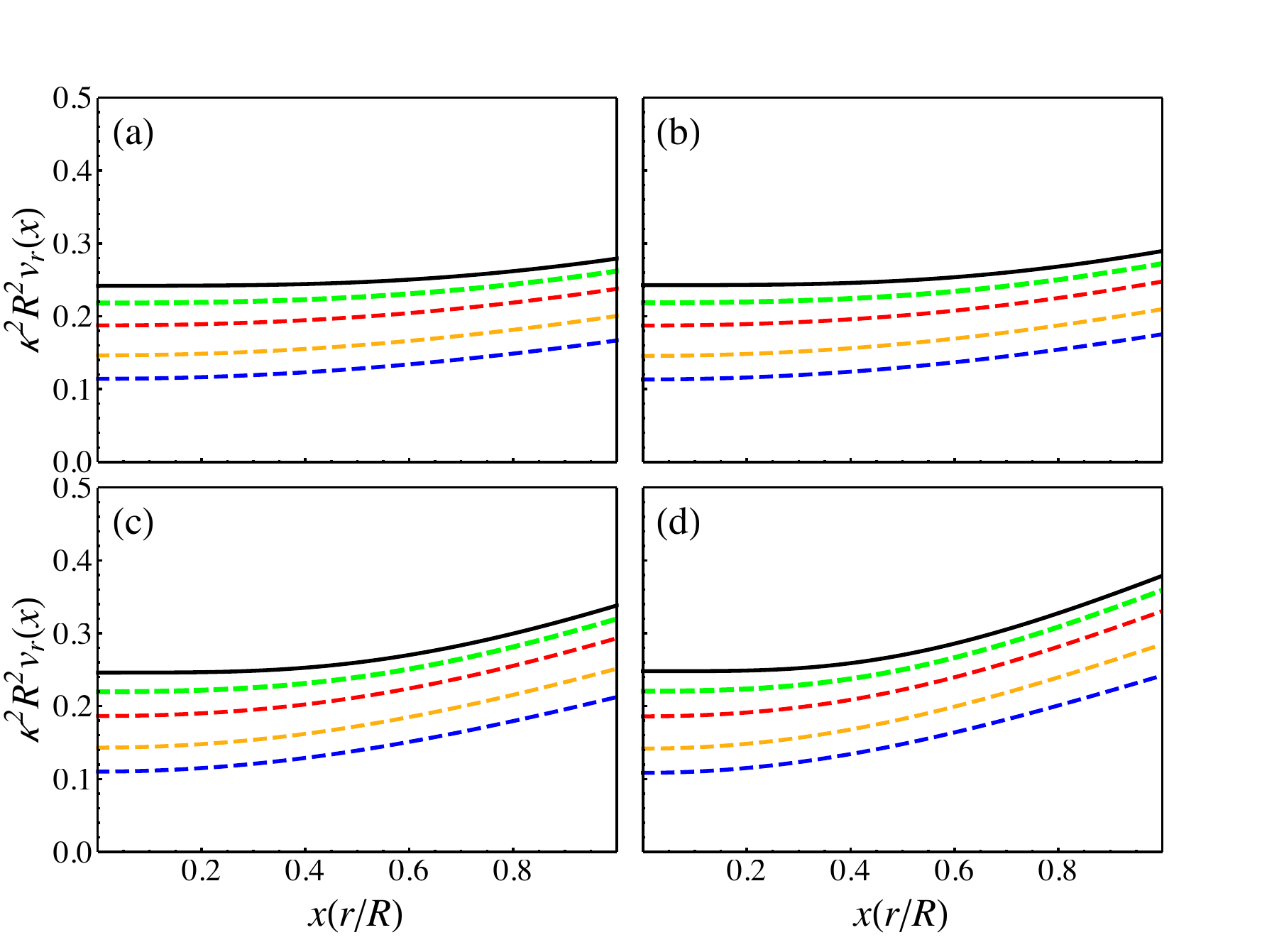}
\caption{\label{vr} 
Radial sound velocity, $v_{r}^{2}$,for 
a) $u=0.13$, b) $u=0.15478$, c) $u=0.2035$, d) $u=0.229365$. $\alpha=0$ (black solid line), $\alpha=0.04$, (green dashed line),
$\alpha=0.1$ (red dashed line), $\alpha=0.2$ (yellow dashed line),
$\alpha=0.3$ (blue dashed line).
}
\end{figure}

\begin{figure}[h!]
\centering
\includegraphics[scale=0.52]{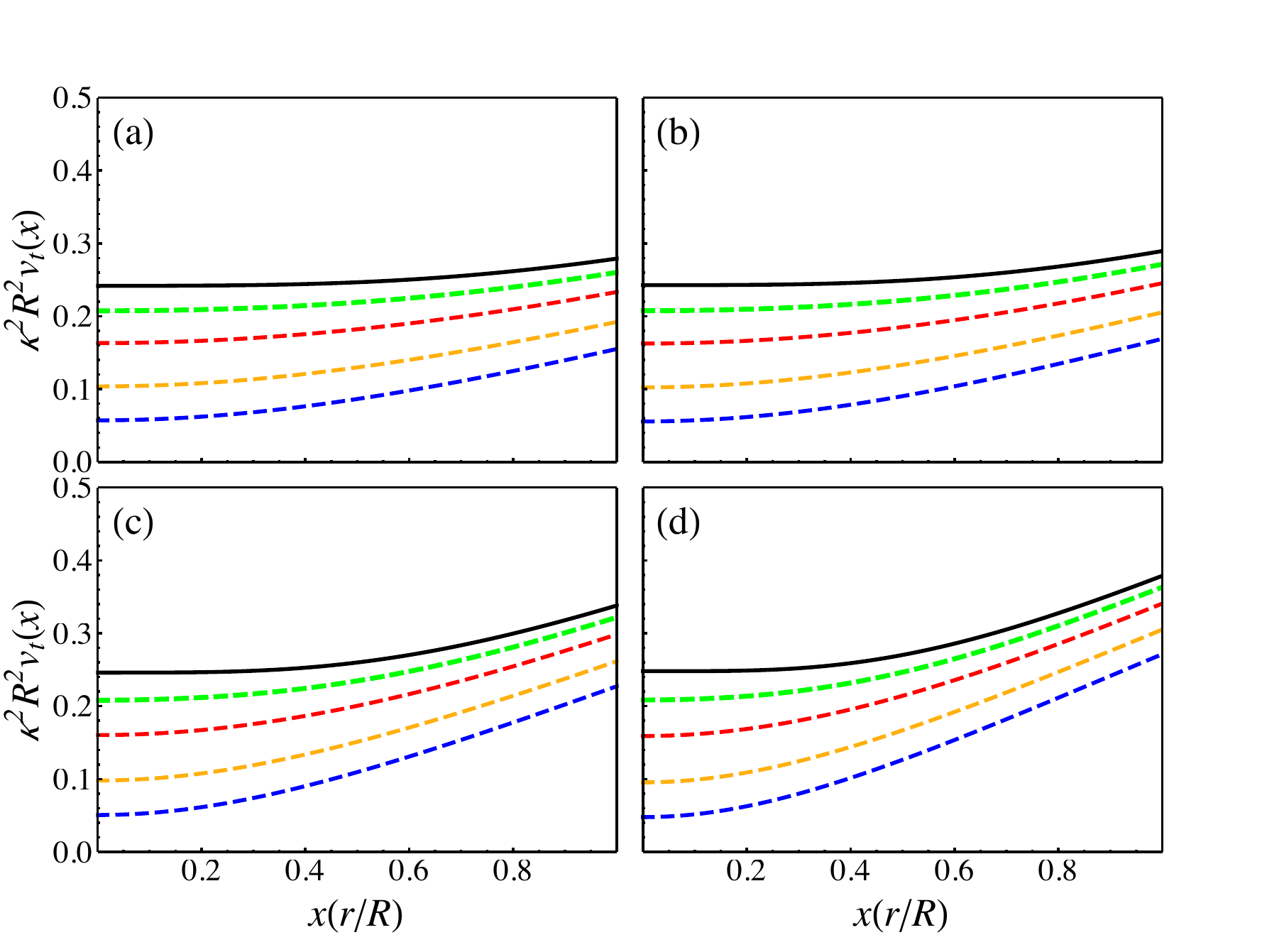}
\caption{\label{vp} 
Traverse sound velocity, $v^{2}_{t}$,  
a) $u=0.13$, b) $u=0.15478$, c) $u=0.2035$, d) $u=0.229365$.  $\alpha=0$ (black solid line), $\alpha=0.04$, (green dashed line),
$\alpha=0.1$ (red dashed line), $\alpha=0.2$ (yellow dashed line),
$\alpha=0.3$ (blue dashed line).
d
}
\end{figure}

The causality condition ensures that
either the radial and tangential sound velocities, $v_{r}=d\tilde{p}_{r}/d\tilde{\rho}$ and $v_{t}=d\tilde{p}_{\perp}/d\tilde{\rho}$ respectively, are less than the speed of light. In figures \ref{vr} and \ref{vp} it is shown that both quantities satisfy the requirement.

\subsection{Adiabatic index}
\begin{figure}[h!]
\centering
\includegraphics[scale=0.52]{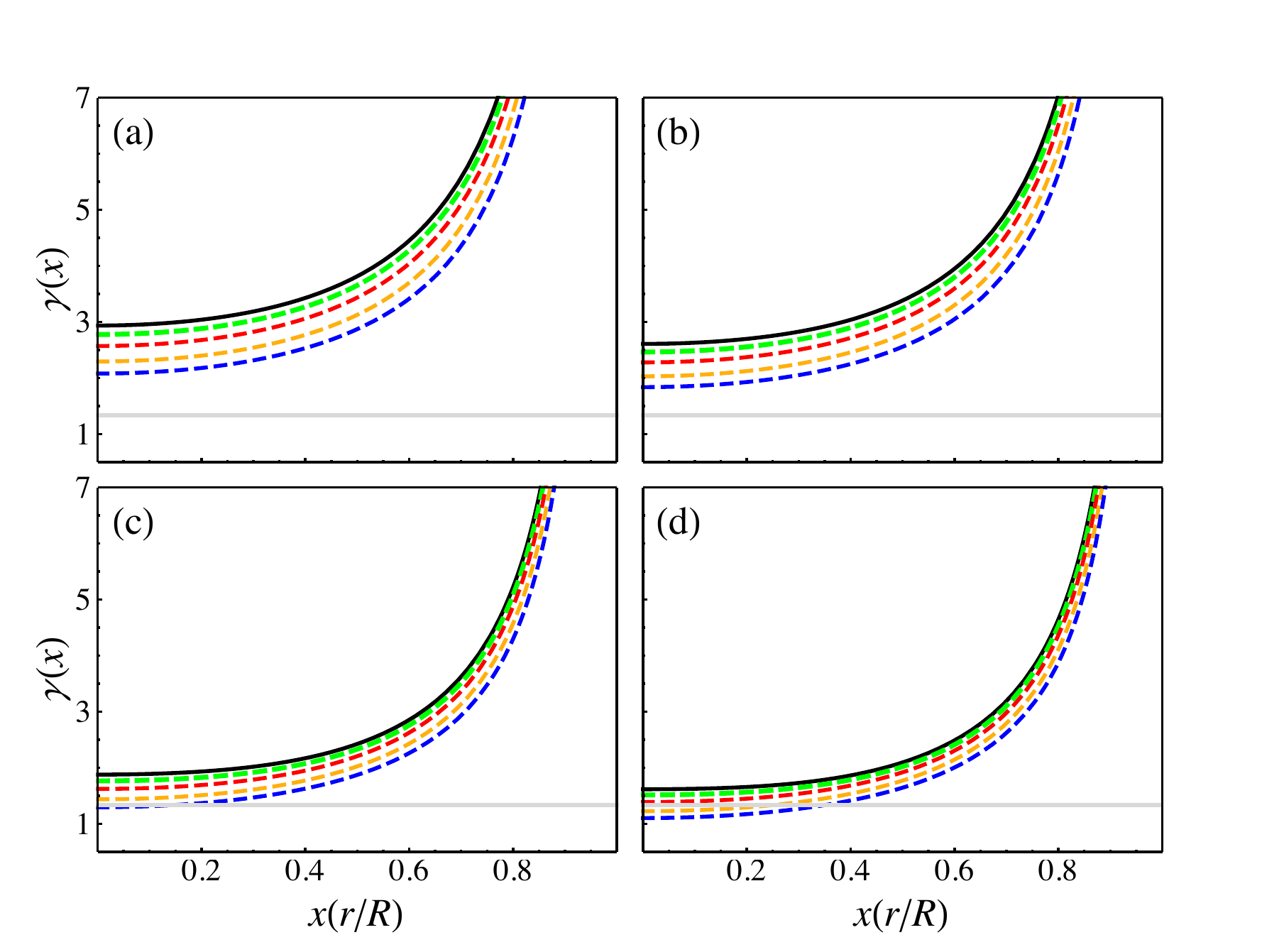}
\caption{\label{ai} 
Adiabatic index, $\gamma$, 
a) $u=0.13$, b) $u=0.15478$, c) $u=0.2035$, d) $u=0.229365$.  $\alpha=0$ (black solid line), $\alpha=0.04$, (green dashed line),
$\alpha=0.1$ (red dashed line), $\alpha=0.2$ (yellow dashed line),
$\alpha=0.3$ (blue dashed line)
}
\end{figure}

As it is well know, the adiabatic index, 
\begin{eqnarray}
\gamma=\frac{\tilde{\rho}+\tilde{p}_{r}}{\tilde{p}_{r}}\frac{d\tilde{p}_{r}}{d\tilde{\rho}},
\end{eqnarray}
allows to connect the relativistic structure of a spherical static object and the equation of state of the interior fluid and serves as a criterion of stability \cite{must}. More precisely, it is said that an interior configuration is stable whenever $\gamma\ge4/3$. Figure \ref{ai} depicts the adiabatic index for different values of $\alpha$. Note that as $\alpha$
increases the system tends to be unstable. Particularly for $u=0.2035$ and $u=0.229365$ the condition $\gamma\ge 4/3$ is violated for $\alpha=0.2$ and $\alpha=0.3$. In this sense, the most stables solutions, are those with smallest compactness parameters which, in our case, correspond to $u=0.13$ and $u=0.15478$. 

\subsection{Convection stability}
\begin{figure}[h!]
\centering
\includegraphics[scale=0.52]{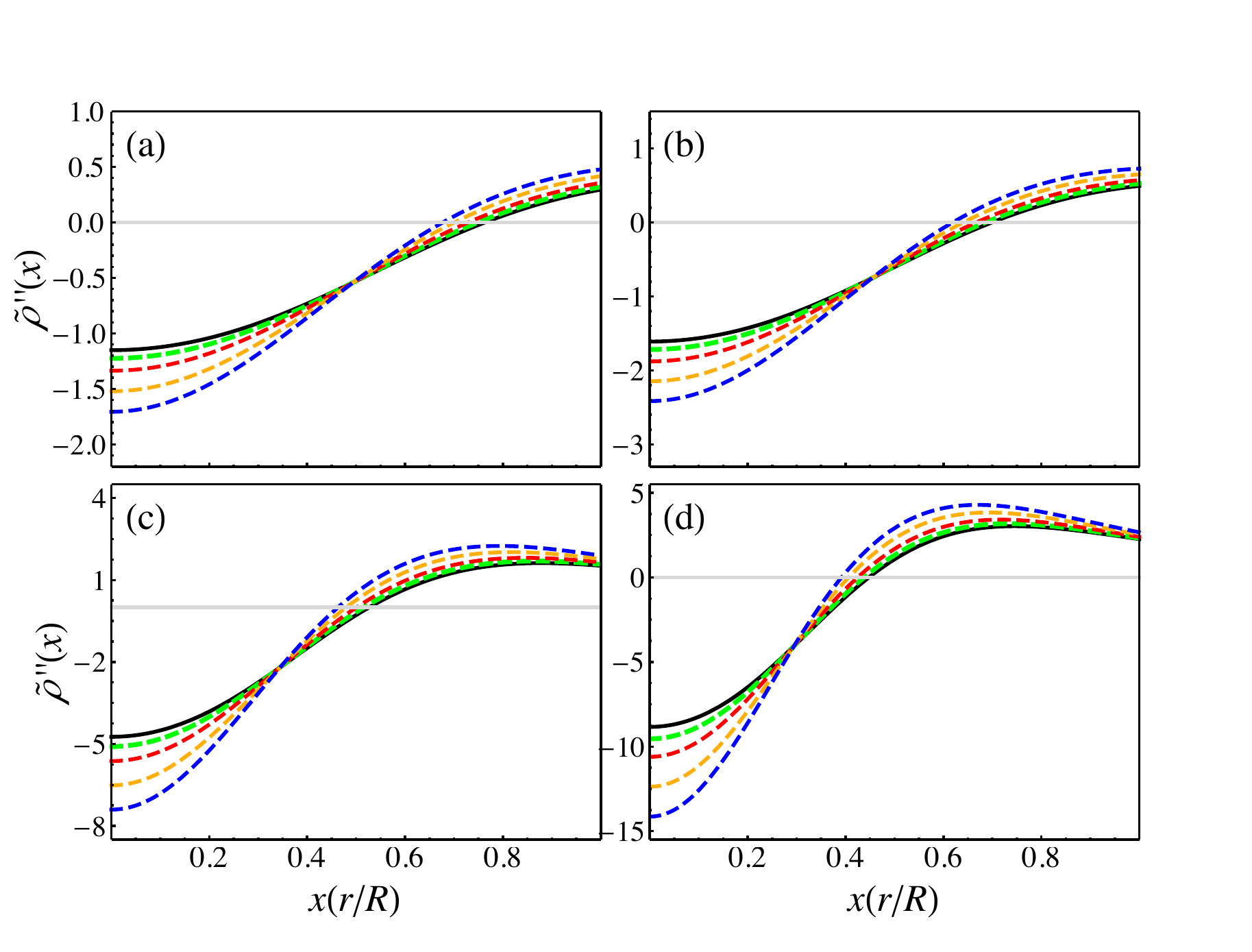}
\caption{\label{cs} 
Second derivative of the density, $\tilde{\rho}$ for 
a) $u=0.13$, b) $u=0.15478$, c) $u=0.2035$, d) $u=0.229365$. $\alpha=0$ (black solid line), $\alpha=0.04$, (green dashed line),
$\alpha=0.1$ (red dashed line), $\alpha=0.2$ (yellow dashed line),
$\alpha=0.3$ (blue dashed line).
}
\end{figure}

A star is stable against convection 
when a fluid element displaced downward floats back to its initial position. As can be demonstrated (see Ref. \cite{nunez}), this occurs whenever
\begin{eqnarray}
\tilde{\rho}''\le 0.
\end{eqnarray}
In figure \ref{cs}
it is shown that all the models under consideration are unstable after undergoing convective motion even in the isotropic case which corresponds to $\alpha=0$

\subsection{Stability against gravitational cracking}
\begin{figure}[h!]
\centering
\includegraphics[scale=0.52]{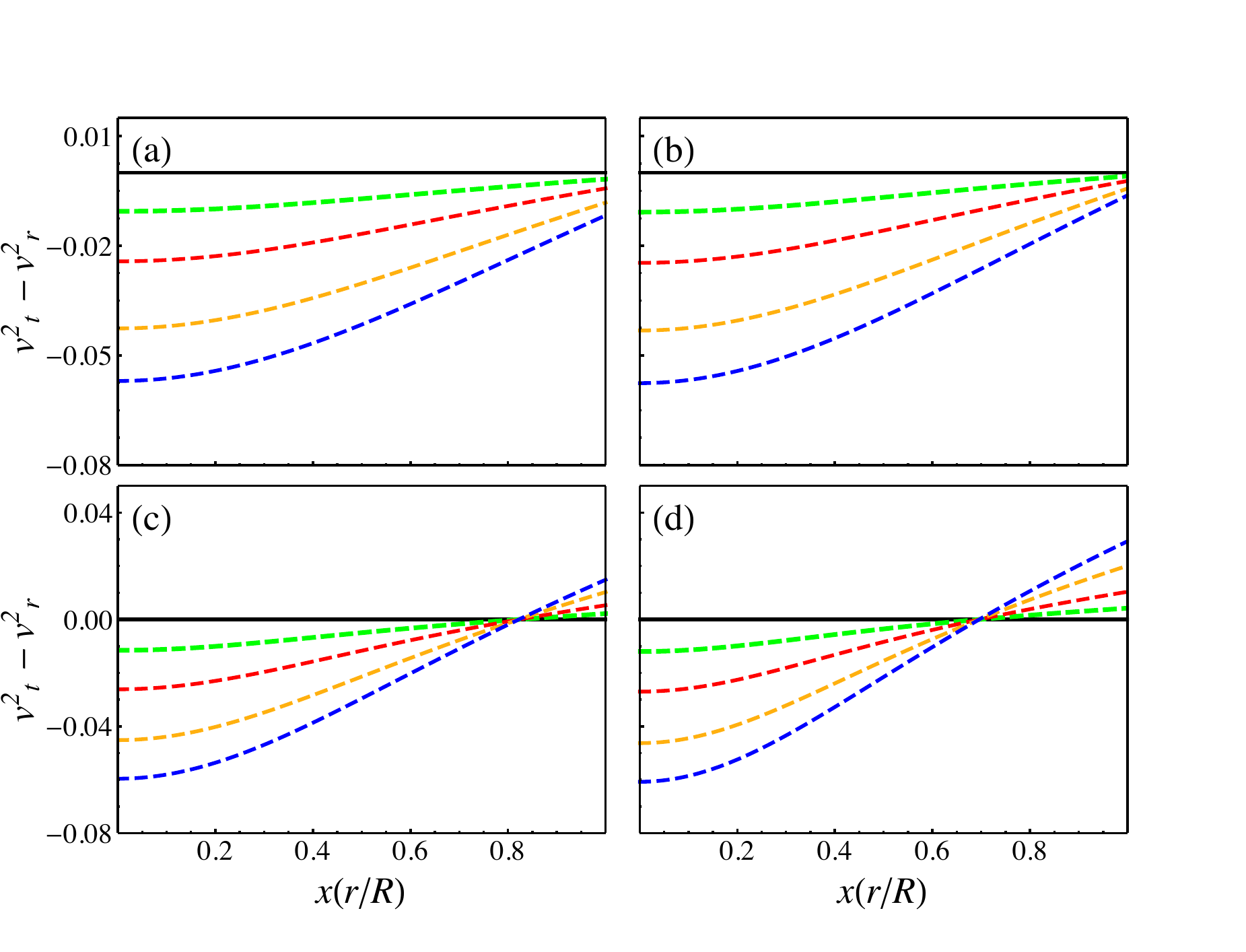}
\caption{\label{crack} 
$\frac{d\tilde{p}_{\perp}}{d\rho}-\frac{d\tilde{p}_{r}}{d\rho}$ for $\alpha=-0.1$, $\tilde{\rho}$
a) $u=0.13$, b) $u=0.15478$, c) $u=0.2035$, d) $u=0.229365$.  $\alpha=0$ (black solid line), $\alpha=0.04$, (green dashed line),
$\alpha=0.1$ (red dashed line), $\alpha=0.2$ (yellow dashed line),
$\alpha=0.3$ (blue dashed line).
}
\end{figure}

The appearance of non--vanishing total radial force with different signs in different regions of the fluid is called gravitational cracking when this radial force is 
directed inward in the inner part of the sphere for all
values of the radial coordinate $r$ between the center
and some value beyond which the force
reverses its direction \cite{lhcrack}. In reference \cite{abreu} it is stated that a simple requirement to avoid gravitational cracking is
\begin{eqnarray}\label{eqc}
-1\le\frac{d\tilde{p}_{\perp}}{d\tilde{\rho}}-
\frac{d\tilde{p}_{r}}{d\tilde{\rho}}\le 0.
\end{eqnarray}
In figure \ref{crack} we show that for $u=0.13$ and $u=0.15478$ the anisotropic solution is stable against cracking for all the values of $\alpha$ considered. In contrast, for $u=0.2035$ and $u=0.229365$
the condition (\ref{eqc}) is clearly not satisfied and the system is unstable against gravitational cracking.

\newpage

\section{Final remarks}\label{remarks}
In this work we have implemented the Minimal Geometric Deformation decoupling method to extend an isotropic model of Neutron Stars to anisotropic domains. More precisely, we studied the model used in \cite{estevez2018} to describes the PSR J0348+0432 compact object but, additionaly, we have considered the models SAX J1808.4-3658, Her X-1 and Cen  X-3, also. Furthermore, we have studied the physical acceptability of all the models for different values of the MGD--parameter, $\alpha$, starting from the isotropic case ($\alpha=0$) and exploring how their behaviour are modified when the anisotropy is induced ($\alpha\ne 0$).
In particular we studied the profiles of density and pressures, energy condition, causality,  and the stability of the solution studying the adiabatic index, the convection and cracking conditions. We found that, for all the values of compactness parameters considered, the density and the pressures satisfy the basic requirements of an interior solution, namely, they reach their maximum value at the center and are monotonously decreasing toward the surface of the star. Additionally, the solution satisfies the dominant and the strong energy conditions and their sound velocities (radial and tangential) are less than the speed of light as expected. We found that, for
compactness parameters corresponding to
SAX J1808.4-3658 and Her X-1 compact objects, the solution is stable according to the adiabatic index criteria for all the values of the decoupling parameter $\alpha$ considered. However, for Cen X-3 and PRS J0348+0432 the adiabatic index is less than $4/3$ for the highest values of $\alpha$ considered and as a consequence, the configuration is unstable in this situations. This situation is similar to the obtained in the study of gravitational cracking, more precisely, the configuration is stable for the most compact object considered and unstable for the other two models. Interestingly, in this case, the instability appears for any value of $\alpha$ we considered. Regarding the convection stability, we found that the solution is unstable for all the models under study for any value of $\alpha$ including the isotropic case which corresponds to $\alpha=0$. In conclusion, we have obtained that according to the adiabatic index and gravitational cracking criterion, the
most stable anisotropic models are those with the smallest compactness parameters, namely SAX J1808.4-3658 and Her X-1.

\end{document}